\documentclass[linembers,trackchanges, twocolumn]{aastex701}

\usepackage{amsmath}
\usepackage{appendix}
\usepackage{ulem}
\usepackage{soul}

\begin{document}

\title{Probing Variations in Earth's Ionosphere Using Pulsars}

\author[orcid=0009-0009-9063-9118]{Yanqing Cai}
\affiliation{State Key Laboratory of Radio Astronomy and Technology, National Astronomical Observatories, CAS, Beijing 100101, China}
\affiliation{University of Chinese Academy of Sciences, CAS, Beijing 100049, China}
\email{caiyq@bao.ac.cn}  

\author[orcid=0000-0002-1381-7859]{Ziwei Wu}
\affiliation{State Key Laboratory of Radio Astronomy and Technology, National Astronomical Observatories, CAS, Beijing 100101, China}
\email[show]{wuzw@bao.ac.cn}  

\author[0000-0001-5105-4058]{Weiwei Zhu}
\affiliation{State Key Laboratory of Radio Astronomy and Technology, National Astronomical Observatories, CAS, Beijing 100101, China}
\affiliation{Institute for Frontier in Astronomy and Astrophysics, Beijing Normal University, Beijing 102206, China}
\email[show]{zhuww@nao.cas.cn}  

\author[0000-0002-4088-896X]{Joris P. W. Verbiest}
\affiliation{Fakult\"at f\"ur Physik, Universit\"at Bielefeld, Postfach 100131, 33501 Bielefeld, Germany}
\email[]{verbiest@physik.uni-bielefeld.de}

\author[0000-0002-5400-6948]{Ningbo Wang}
\affiliation{Aerospace Information Research Institute (AIR), Chinese Academy of Sciences (CAS), Beijing 100094, China}
\email{wangningbo@aoe.ac.cn}

\author[]{Moochickal Ambalappat Krishnakumar}
\affiliation{National Centre for Radio Astrophysics, Tata Institute of Fundamental Research, Pune 411007, Maharashtra, India}
\email{kk.ambalappat@gmail.com}

\author[0000-0003-4634-5453]{Lars K\"unkel}
\affiliation{Department of Physics, McGill University, 3600 rue University, Montr\'eal, QC H3A 2T8, Canada}
\affiliation{Trottier Space Institute, McGill University, 3550 rue University, Montr\'eal, QC H3A 2A7, Canada}
\email{lkuenkel@phas.ubc.ca}

\author[]{J\"orn K\"unsemöller}
\affiliation{Fakult\"at f\"ur Physik, Universit\"at Bielefeld, Postfach 100131, 33501 Bielefeld, Germany}
\email{jkuensem@physik.uni-bielefeld.de}

\author[0000-0001-9986-9360]{Yulan Liu}
\affiliation{CAS Key Laboratory of FAST, National Astronomical Observatories, CAS, Beijing 100101, China}
\affiliation{Guizhou Radio Astronomical Observatory, Guizhou University, Guiyang 550001, China}
\email{liuyl@bao.ac.cn}

\author[0000-0002-6955-8040]{Nataliya Porayko}
\affiliation{Max-Planck-Institut f\"ur Radioastronomie, Auf dem H\"ugel 69, D-53121 Bonn, Germany}
\email{porayko.nataliya@gmail.com}

\author[0000-0002-8452-4834]{Golam M. Shaifullah}
\affiliation{Dipartimento di Fisica ``G. Occhialini'', Universit\`a di Milano-Bicocca, Piazza della Scienza 3, 20126 Milano, Italy}
\affiliation{INFN, Sezione di Milano-Bicocca, Piazza della Scienza 3, I-20126 Milano, Italy}
\email{golam.shaifullah@UNIMIB.IT}

\author[0000-0001-6651-4811]{Caterina Tiburzi}
\affiliation{INAF - Osservatorio Astronomico di Cagliari, via della Scienza 5, 09047 Selargius (CA), Italy}
\email{1984cat.ti@gmail.com}

\author[]{Marcus Brüggen} 
\affiliation{Hamburger Sternwarte, University of Hamburg, Gojenbergsweg 112, 21029 Hamburg, Germany}
\email{mbrueggen@hs.uni-hamburg.de}

\author[0000-0001-8206-5956]{Ralf-Jürgen Dettmar} 
\affiliation{Ruhr-Universität Bochum, Fakultät für Physik und Astronomie, Astronomisches Institut, 44780 Bochum, Germany}
\email{dettmar@astro.rub.de}

\author[]{Ziyao Fang}
\affiliation{State Key Laboratory of Radio Astronomy and Technology, National Astronomical Observatories, CAS, Beijing 100101, China}
\email{fangzy@bao.ac.cn}

\author[]{Qiuyang Fu}
\affiliation{State Key Laboratory of Radio Astronomy and Technology, National Astronomical Observatories, CAS, Beijing 100101, China}
\affiliation{University of Chinese Academy of Sciences, CAS, Beijing 100049, China}
\email{fuqy@bao.ac.cn}

\author[]{Jiawei Jin}
\affiliation{State Key Laboratory of Radio Astronomy and Technology, National Astronomical Observatories, CAS, Beijing 100101, China}
\affiliation{University of Chinese Academy of Sciences, CAS, Beijing 100049, China}
\email{jinjw@bao.ac.cn}

\author[]{Caisong Liu}
\affiliation{State Key Laboratory of Radio Astronomy and Technology, National Astronomical Observatories, CAS, Beijing 100101, China}
\affiliation{University of Chinese Academy of Sciences, CAS, Beijing 100049, China}
\email{liucs@bao.ac.cn}

\author[0000-0002-2885-568X]{Lingqi Meng}
\affiliation{State Key Laboratory of Radio Astronomy and Technology, National Astronomical Observatories, CAS, Beijing 100101, China}
\email{menglingqi@nao.cas.cn}

\author[0000-0003-1185-8937]{Xueli Miao}
\affiliation{State Key Laboratory of Radio Astronomy and Technology, National Astronomical Observatories, CAS, Beijing 100101, China}
\email{xlmiao@bao.ac.cn}

\author[0000-0001-8065-4191]{Jiarui Niu}
\affiliation{State Key Laboratory of Radio Astronomy and Technology, National Astronomical Observatories, CAS, Beijing 100101, China}
\email{niujiarui@nao.cas.cn}

\author[0000-0003-2413-0881]{Dominik J. Schwarz} 
\affiliation{Fakult\"at f\"ur Physik, Universit\"at Bielefeld, Postfach 100131, 33501 Bielefeld, Germany}
\email{dschwarz@physik.uni-bielefeld.de}

\author[0000-0001-8583-8619]{Christian Vocks} 
\affiliation{Leibniz-Institut f\"ur Astrophysik Potsdam (AIP), An der Sternwarte 16, 14482 Potsdam, Germany}
\email{cvocks@aip.de}

\author[0000-0001-8018-1830]{Mengyao Xue}
\affiliation{State Key Laboratory of Radio Astronomy and Technology, National Astronomical Observatories, CAS, Beijing 100101, China}
\email{mengyaoxue@nao.cas.cn}

\author[0000-0003-1874-0800]{Mao Yuan}
\affiliation{National Space Science Center, CAS, Beijing 100190, China}
\email{yuanmao@nssc.ac.cn}

\author[0000-0003-4415-2148]{Youling Yue}
\affiliation{State Key Laboratory of Radio Astronomy and Technology, National Astronomical Observatories, CAS, Beijing 100101, China}
\email{ylyue@bao.ac.cn}

\author[]{Zhen Zhang}
\affiliation{Institute for Frontier in Astronomy and Astrophysics, Beijing Normal University, Beijing 102206, China}
\email{zhangz@bao.ac.cn}

\begin{abstract}
We present high cadence 10-minute rotation measure (RM) monitoring of PSR~J0814+7429 using the
LOw-Frequency ARray, aiming to probe ionospheric variability along the pulsar line of sight (LoS).
By separating the ionospheric contribution from the observed RM, we quantitatively reconstruct the diurnal variation of the ionospheric electron density along the pulsar LoS based on the World Magnetic Model. 
The derived variations exhibit clear solar-driven modulation, including the ionospheric noontime bite-out phenomenon, and show good agreement with the LoS total electron content reconstructed from independent global vertical total electron content maps.
These results demonstrate the feasibility of using pulsars as probes of temporal variations in the electron density of the Earth's ionosphere.
\end{abstract}

\keywords{\uat{Pulsars}{1306} -- \uat{Earth ionosphere}{860}}

\section{Introduction} 
Radio pulsars are highly magnetized and rapidly rotating neutron stars, which emit radio waves in two highly collimated beams from their magnetic poles \citep{pai67}.
As pulsar signals propagate through the interstellar medium (ISM), they are usually affected in a number of ways, including frequency dispersion \citep{lk12}, Faraday rotation \citep{man71}, scintillation \citep{sch68}, and scattering \citep{ric77}.
Among these propagation effects, the Faraday rotation (FR) is induced by the magnetised plasma along the line of sight (LoS) and described by the rotation measure (RM):
\begin{equation}\label{equ1}
\begin{split}
         \Delta\psi & = \mathrm{RM}\, \lambda^{2},\\
         \mathrm{RM} & = 0.81 \int_{\text{source}}^{\text{observer}} 
n_{e}\,\mathbf{B}\cdot d\mathbf{r} \; [\mathrm{rad}\,\mathrm{m}^{-2}], 
\end{split}
\end{equation}
where $\Delta \psi$ is the linear polarization position angle of the pulsar radiation, $\lambda$ the
observing wavelength, $n_{e}$ the interstellar electron density in units of $\rm cm^{-3}$,
$\mathbf{B}$ the magnetic field vector in microgauss and $d\mathbf{r}$ an elemental vector along the LoS toward us in units of pc \citep{bb05}.
The integral extends along the full line of sight from the source to the observer, of which the major contribution to the FR is given by the magneto-ionic ISM.
Furthermore, owing to the relatively strong geomagnetic field and the high electron density in the
Earth’s ionosphere, this region also contributes significantly to the total FR along the LoS to the pulsar \citep{ymh+11}.
Consequently, when the pulsar signal propagates through Earth’s ionosphere, temporal variations, like
diurnal variations, in the ionospheric plasma will be imprinted in the observed signal \citep{ssh+13}.

Therefore, the signals from pulsars may provide valuable information about the Earth’s ionosphere and thereby serve as an effective probe of ionospheric variability.
The results of such pulsar–ionosphere studies have been carried out using a variety of observational techniques, including polarization measurements \citep[e.g.,][]{usm+13,ssh+13,pnt+19}, radio interferometry and scintillation analyses \citep[e.g.,][]{zya+20,bp24}.
However, most previous studies have mainly focused on validating ionospheric FR corrections,  or comparing different ionospheric models.
While densely sampled, nearly continuous pulsar observations that follow the diurnal evolution of the ionosphere remain limited.
Such measurements are important for testing pulsars as stable LoS probes of ionospheric variability.
Motivated by this, in this letter we use more than 30 days of polarization observations of PSR J0814$+$7429 to investigate the temporal evolution of the ionospheric FR and to further explore the temporal variations in the electron density of the Earth's ionosphere along the LoS.
In Section~\ref{sec2}, we describe our observations and data processing procedures.
The results are given in Section~\ref{sec3}.
Section~\ref{sec4} presents the discussions and conclusions.

\section{Observations and Data Analysis}\label{sec2}
The data in this letter observed using the LOw Frequency ARray \citep[LOFAR;][]{vwg+13} and the observations were taken with stations of the German LOng Wavelength
(GLOW\footnote{\url{https://www.glowconsortium.de/}}) array, which conducts observing campaigns of pulsars at low frequencies ($\sim$100 MHz$-$200 MHz) weekly \citep{tvs+19,dvt+20}.

The data sets of PSR~J0814$+$7429 were observed in 2017, with three German LOFAR stations: DE604 in Bornim, DE605 in J\"ulich, and DE609 in Norderstedt.
Each observation was operated in stand-alone mode, but with an identical observing setup.
Among the three observing stations, DE604 and DE609 provide continuous observations throughout the entire day, whereas DE605 only offers a few hours of continuous observations per day.
The observations have 195~kHz frequency resolution with 366 channels recorded using the LUMP
Software\footnote{\url{https://github.com/AHorneffer/lump-lofar-und-mpifr-pulsare}}, over a
frequency range from $\sim$120 MHz to $\sim$190 MHz.
Data pre-processing, including de-dispersed, polarization calibration, and basic radio-frequency
interference (RFI) mitigation, was applied as described in detail elsewhere \citep{dvt+20,tvs+19}.
For the work presented here, further processing was carried out as follows: all observations
underwent additional excision of RFI using a modified version\footnote{\url{https://github.com/larskuenkel/iterative_cleaner}} of the SURGICAL algorithm from the COASTGUARD package \citep{lkg+16}.

A total of 32 days of observations were selected for analysis, of which only a representative subset is shown here for clarity (see Section \ref{sec3}), as the remaining results exhibit similar behavior. 
Except for March 10, March 17, and October 8, which had partial coverage, data collection on all other days encompassed full 24-hour cycles.
Each day's observation consisted of multiple consecutive data sets, separated by $\sim$3-minute gaps.
We used the PSRADD tool in the PSRCHIVE \citep{vdo12} software package to add all consecutive data sets together for each day, and then the added observation was split into 10-minute segments. 
Finally, we used the RMFIT tool in the PSRCHIVE software package to calculate the observed RM for each segment, and removed outliers with excessively large uncertainties or points with poorly fitting results based on visual inspection. 
A comparison between the RM values obtained with RMFIT and those calculated using the RM Synthesis technique \citep{nsk+15,pnt+19} shows good agreement, with both methods yielding RM values close to –13 rad m$^{-2}$.
The ionospheric RMs along the LoS were estimated using the publicly available RMEXTRACT python package \citep{mev+18} with the global ionosphere map (GIP) from the Center for Orbit Determination in Europe (CODE\footnote{\url{ftp://ftp.unibe.ch/aiub/CODE/}}) and the world magnetic model \citep[WMM\footnote{\url{https://www.ngdc.noaa.gov/geomag/WMM/}},][]{wmm2020}.
Among the tested GIM products, the CODE maps showed relatively good agreement with the observed RM variations, particularly in reproducing the overall temporal behavior of the measurements, and were therefore adopted in this work.

\begin{figure*}
    \centering
    \includegraphics[width=\linewidth]{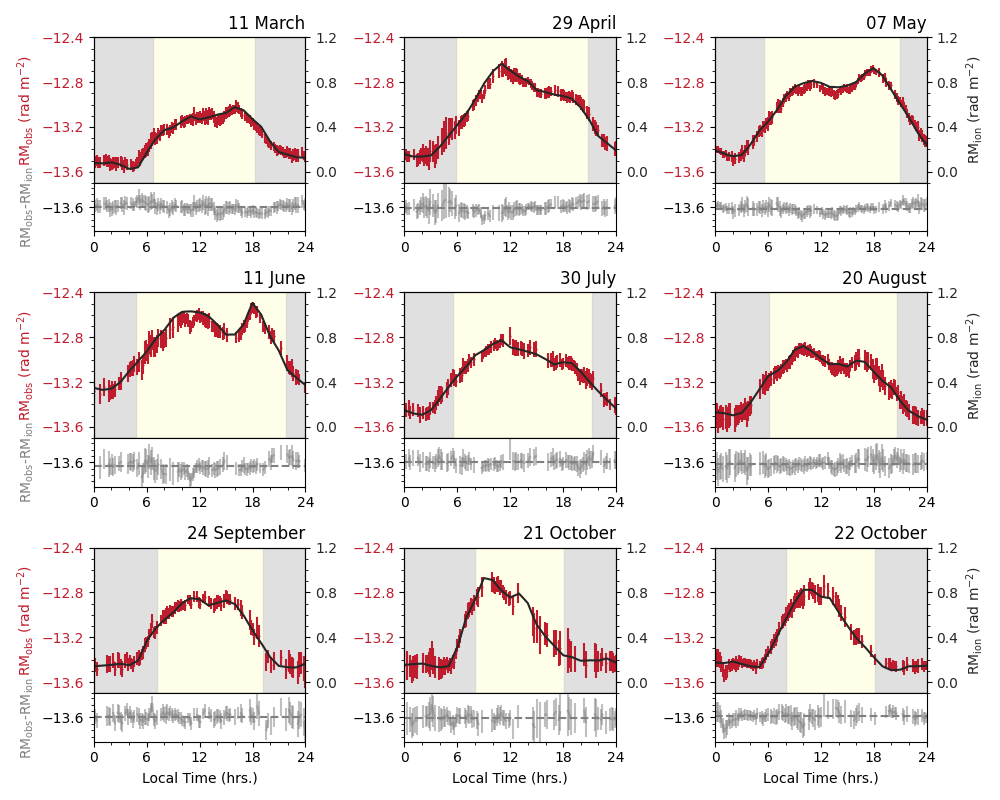}
    \caption{The time series of RMs as a function of local time during a day for nine
      observations made in 2017. For the top panel in each plot, the red points show the RM of PSR~J0814$+$7429 (left axis label, $\rm RM_{obs}$) and the black line shows the ionospheric RM (right
      axis label, $\rm RM_{ion}$) along the LoS to this pulsar (as derived by the RMEXTRACT software package, based on the CODE+WMM). The bottom panel in each plot, the gray points represent the residuals between the $\rm RM_{obs}$ and $\rm RM_{ion}$, and the gray horizontal line is the mean value of these residuals. The observation time is labeled in the upper-right of each plot. From left to right in each plot, the first and second boundaries between the light-gray and yellow shaded regions mark the sunrise and sunset times, respectively. That is, the light-gray shaded region mainly corresponds to nighttime, and the light-yellow shaded region to daytime. The ISM RM is approximately $\rm -13.60\, rad\,m^{-2}$ (see Section~\ref{probied}). Except for the observation on May 7, which was conducted using the DE604 station, the other observations were all conducted using the DE609 station. }
    \label{fig:rm_pulsar_ionosphere}
\end{figure*}

\section{Results}\label{sec3}

\subsection{The Diurnal Variations in the Observed RMs}
In Figure \ref{fig:rm_pulsar_ionosphere}, we demonstrated nine representative RM series as a function of the station's local time.
It can be clearly seen that the RM time series shows pronounced diurnal variations.
The estimated RM values of the ionosphere (hereafter $\rm RM_{ion}$) and the observed RM values of RSR J0814$+$7429 (hereafter $\rm RM_{obs}$) agree well. 
The results in Figure \ref{fig:rm_pulsar_ionosphere} show that the amplitude of the daily variations in $\rm RM$ is approximately $\rm 0.5 \sim 1\, rad\,m^{-2}$.
Related research \citep{pnt+19} shows that stochastic fluctuations in RM due to turbulence in the ISM are expected to be of the order $\rm 10^{-5}-10^{-4}\, rad\,m^{-2}$ for observing campaigns of $\rm \sim 1-5\,yr$, which is approximately four to five orders of magnitude lower than the RM variations observed in this work. 
Furthermore, PSR~J0814$+$7429 is not in a binary system, so no additional RM contribution from a companion’s magnetic field is expected \citep[e.g.,][]{lbr+23}.
Therefore, the amplitude of the daily variations in $\rm RM_{obs}$ is mainly related to the changes in the ionosphere, which forms the basis of using pulsar signals to probe ionospheric variability.

\begin{figure*}
    \centering
    \includegraphics[width=\linewidth]{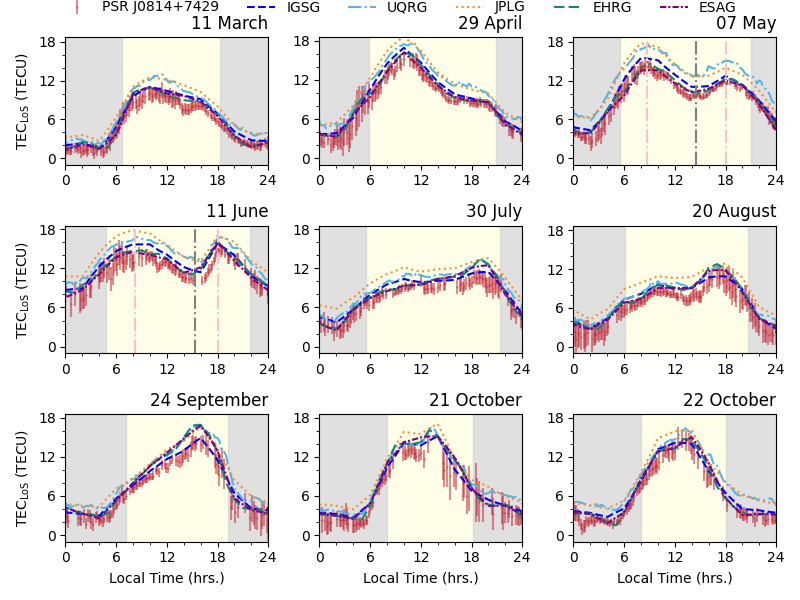}
    \caption{The time series of electron density within the ionosphere along the LoS ($\rm TEC_{LoS}$) as a function of local time during a day for nine representative observations, where the red points represent the $\rm TEC_{LoS}$ calculated by the pulsar signal (labelled as PSR~J0814$+$7429), different colors and line styles (labeled on the top of the Figure) correspond to $\rm TEC_{LoS}$ obtained from different global ionospheric maps. The meanings of the gray area, yellow area, and the boundaries between them are consistent with Figure \ref{fig:rm_pulsar_ionosphere}. In the corresponding subplots on May 7 and June 11, the vertical pink dotted lines mark the peak positions before and after the noontime bite-out, while the vertical gray dotted lines indicate the minimum value during the bite-out (see Section \ref{bite-out}).
    The observation time is labeled in the upper-right of each plot. 
    Selected unit: $\rm 1\,TECU=1\times10^{16}\,electron/m^{2}$.}
    \label{fig:TEC_pulsar_code_igsg}
\end{figure*}

\subsection{Probing the Ionospheric Electron Density}\label{probied}
To extract the relatively pure ionospheric RM variations along the LoS to PSR~J0814$+$7429 from the observations, it is necessary to calculate the ISM's RM along this LoS.
We calculated the ISM's RM along the LoS to PSR~J0814$+$7429 using the 32 days of observations.
We first compute the daily mean residuals ($\rm RM_{obs} - \rm RM_{ion}$) between $\rm RM_{obs}$ and $\rm RM_{ion}$ to estimate the LoS ISM RM for each day.
A Spearman rank correlation test is then applied to assess the presence of any systematic trend in the resulting 32 daily RM values, and no significant trend is detected. 
We finally take the mean of these values to obtain the ISM RM, yielding a value of $\rm -13.60\pm 0.06\, rad\,m^{-2}$.
Then, it is convenient for us to isolate the ionospheric RM from the $\rm RM_{obs}$ under the assumption
of a constant ISM RM.
The ionospheric RM variations along the LoS, hereafter denoted as $\rm \phi_{obs-ISM}$, are derived by subtracting the final ISM RM from the $\rm RM_{obs}$.
For convenience in calculation, we will treat $\rm \phi_{obs-ISM}$ as equivalent to the value at the ionospheric piercing point (IPP).

Based on the assumption of a thin spherical shell surrounding the Earth with a fixed $\rm H_{eff}$ \citep[e.g.,][]{ssh+13}, and in conjunction with Equation \ref{equ1}, the ionospheric RM is expressed as:
\begin{equation}\label{equ2}
    \rm \phi_{ion}=2.6\times10^{-17}TEC_{LoS}B_{LoS}\, rad\, m^{-2},
\end{equation}
where the units of the $\rm TEC_{LoS}$ and $\rm B_{LoS}$ are $\rm m^{-2}$ and gauss, respectively.
By equating $\rm \phi_{ion}$ with $\rm \phi_{obs-ISM}$ in Equation \ref{equ2}, the value of $\rm TEC_{LoS}$ can be derived as follows:
\begin{equation}\label{equ3}
    \rm TEC_{LoS}=1\times10^{17}\frac{\phi_{obs-ISM}}{2.6 \times B_{LoS}}\, m^{-2}.
\end{equation}
The value of the $\rm B_{LoS}$ was extracted from the WMM using the RMEXTRACT python code at the effective height of the ionosphere.
We also performed linear interpolation on the $\rm B_{LoS}$ to match the calculations.
We adopted a conservative WMM uncertainty of $\rm B_{LoS}$ $\rm \sigma_{B_{LoS}}\sim 150\ nT$ for $\rm B_{LoS}$, based on the WMM error estimates reported by \cite{nfc+25}.
Then, the uncertainty of the derived $\rm TEC_{LoS}$ was estimated by propagating both the $\rm \phi_{obs-ISM}$ uncertainty and the uncertainty of $\rm B_{LoS}$.

We presented nine days of the $\rm TEC_{LoS}$ results for display in Figure \ref{fig:TEC_pulsar_code_igsg}, where the red points represent the $\rm TEC_{LoS}$ calculated by the pulsar signal (labelled as PSR J0814+7429) through Equation \ref{equ3}, and different colors and line styles correspond to $\rm TEC_{LoS}$ values derived from different GIMs.
These GIM products, including IGSG, UQRG, JPLG, EHRG, and ESAG, are publicly accessible through NASA's website\footnote{\url{cddis.nasa.gov/archive/gnss/products/ionex/}}.
As shown in each plot of Figure \ref{fig:TEC_pulsar_code_igsg}, these $\rm TEC_{LoS}$ time series exhibit similar temporal variations, with good agreement during a day. 
The discrepancy between the $\rm TEC_{LoS}$ derived from several GIMs and that inferred from pulsar observations can reach up to $\sim$1--2 TECU ($1\,\mathrm{TECU}=10^{16}\,\mathrm{electrons\,m^{-2}}$), with the GIM-based estimates generally being higher, consistent with a similar offset reported in a recent study \citep{pbg+26}, which used the Moon as a polarization calibrator for absolute polarimetry to correct for ionospheric Faraday rotation.
Such a discrepancy is also inevitable in ionospheric studies, as GIM products provided by different analysis centers often exhibit inter-center differences of several TECU \citep[e.g.,][]{zhl21}.
Each analysis center employs a specific method to generate its respective GIM product, resulting in GIMs that may vary from center to center \citep[e.g.,][]{Liu2026}.
In this context, we consider our results to lie within the typical uncertainty range of current ionospheric GIM products.
The remaining results, not shown here, are similar to those presented in Figure \ref{fig:TEC_pulsar_code_igsg}.
These results demonstrate the feasibility of using the pulsars as probes to detect variations in the Earth’s ionospheric electron
density.

\begin{figure*}
    \centering
    \includegraphics[width=\linewidth]{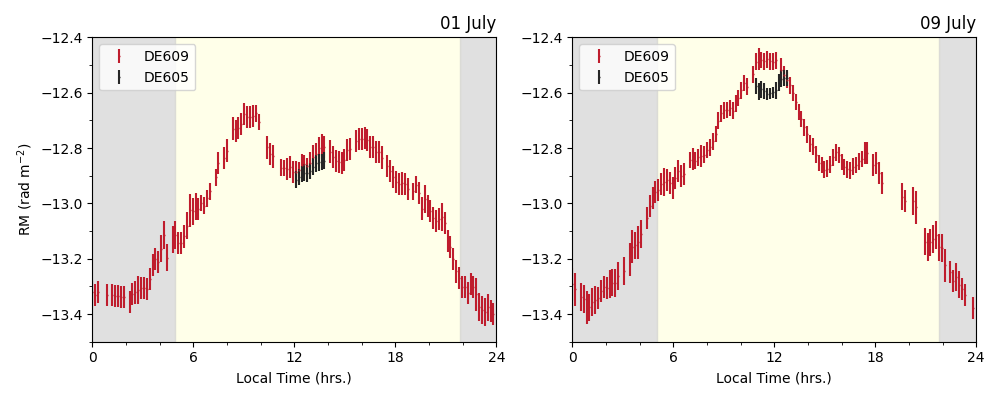}
    \caption{The time series of RMs as a function of local time during a day, where the left and right plots correspond to the observation results from 01 July and 09 July, respectively. The red and black points in each plot were observed using the station DE609 and station DE605. The meanings of the gray area, yellow area, and the boundaries between them are consistent with Figure \ref{fig:rm_pulsar_ionosphere}.}
    \label{fig:de605_de609}
\end{figure*}

\subsection{The Ionospheric Noontime Bite-outs}\label{bite-out}
The phenomenon of the ionospheric noontime bite-out manifests as an anomalous reduction in electron density around midday, characterized by two electron density maxima separated by a pronounced minimum value \citep[e.g.,][]{app+46,lia+47}.
{Based on the method of judging noontime bite-out \citep{hzw+25}:
\begin{equation}
    \rm I_{bite}=(P_{TEC_{LoS}} - V_{TEC_{LoS}})/P_{TEC_{LoS}}>5\%,
\end{equation}
where $\rm P_{TEC_{LoS}}$ corresponds to the $\rm TEC_{LoS}$ peak values before and after the noontime bite-out event, and $\rm V_{TEC_{LoS}}$ represents the minimum $\rm TEC_{LoS}$ value within the bite-out valley.
To ensure the reliability of the identified noontime bite-out structures, both the pre-bite-out and post-bite-out $\rm TEC_{LoS}$ peaks are required to exceed the valley $\rm TEC_{LoS}$ by more than 5\% of their respective peak values.
We identified the results of $\rm TEC_{LoS}$ in our work to check for the presence of noontime bite-outs within them.
To improve the accuracy of identification, only when all the $\rm TEC_{LoS}$ time series from pulsar signal and GIMs (see Figure \ref{fig:TEC_pulsar_code_igsg}) exhibit the noontime bite-outs phenomenon simultaneously can we consider that this phenomenon occurred on that day.
Among our 32 days of observations, we identified two days exhibiting a clear noontime bite-outs phenomenon, shown in Figure \ref{fig:TEC_pulsar_code_igsg} (May 07 and June 11), where the vertical pink dotted lines mark the peak positions before and after the noontime bite-out, while the vertical gray dotted lines indicate the minimum value during the bite-out.
Since the noontime bite-out features differ slightly among these $\rm TEC_{LoS}$ time series, the dotted lines indicate the mean locations of the pre-bite-out peak, post-bite-out peak, and the minimum between them derived from all these datasets.
The minimum values of these two noontime bite-outs events occurred on average at 14:00 local time, and the mean duration is about 8 hours.
These are generally similar to previous results \citep[e.g.,][]{lgr+14,hzw+25}.
Usually, the mechanisms behind noontime bite-outs vary across different latitudes \citep[e.g.,][]{jwy+23}, future pulsar observations could help study this phenomenon and better understand ionospheric dynamics.

\subsection{The Impact of Solar Activity on the Ionosphere}
Within the selected dataset, simultaneous observations of PSR J0814+7429 were conducted by the DE605 and DE609 stations on 11 observing days, with each joint observing session lasting for several hours.
In Figure \ref{fig:de605_de609}, we show two days of the RM series for this pulsar, where the red and black points are observed using the stations DE609 and DE605, respectively. 
Accounting for the time difference arising from the longitudinal difference between the two stations, we aligned the two RM sequences by shifting the RM series from DE605 to the left by approximately 15 minutes.
Of the total 11 continuous simultaneous observations, the left plot in Figure~\ref{fig:de605_de609} (observed on 01 July) shows a high degree of consistency between the RM series from the two stations, which is representative of the agreement seen during the other 10 observations; in contrast, the agreement is considerably weaker in the right plot (observed on 09 July), which represents the sole exception.
A relevant report\footnote{\url{https://earthsky.org/space/july-7-8-9-2017-sunspot-geomagnetic-storm-auroras/}} shows that Solar storms had been forecast for July 9, 2017, when a solar wind stream was projected to impact Earth's magnetic field.
Such space weather events could modify the ionospheric electron density distribution and geomagnetic environment, thereby introducing additional variability into FR measurements and affecting the observed RM values.
Therefore, we speculate that the difference in RM observations between the two stations on 9 July may be attributed to the influence of solar activity on the ionosphere. 

From the results presented above, as the pulsar signal propagates through the ionosphere disturbed by solar activity or other events, the high–time–resolution RM measurements of pulsars therefore provide a potential advantage for ionospheric studies, enabling the detection of short-timescale ionospheric dynamics and improving our understanding of the near-Earth space environment.
Similar results have also recently been discussed in \cite{khz+25} and pulsars are  expected to demonstrate  potential as supplementary probes for space weather studies.

\section{Discussions and conclusions}\label{sec4}
Currently, the GNSS is one of the primary technical methods for detecting the Earth's ionosphere \citep{mwy+98}, which provides a rich data resource for global ionospheric research and applications \citep{fel03}.
Accurately capturing the variation of the Earth’s ionosphere is of great importance, as it not only facilitates improved satellite navigation, communication, and space weather forecasting, but also enhances our understanding of the near-Earth space environment \citep[e.g.,][]{sja+24,lzw+25}.
In recent years, many efforts have been made to improve the accuracy of GNSS-based ionospheric monitoring.
For example, \cite{yl+21} proposed utilizing the existing railway network by deploying GNSS receivers on high-speed train platforms for ionospheric observation. 
\cite{sja+24} proposed a novel approach that leverages millions of Android smartphones as a distributed sensor network, reconstructing the global ionosphere using dual-frequency GNSS measurements.
These novel methods show that the coordinated use of multiple data sources in the future may provide a more powerful approach for probing ionospheric variability.

Although this letter provides $\rm TEC_{LoS}$ measurements only along the LoS toward a single pulsar, our results demonstrate that pulsar observations could serve as an effective complementary probe to GNSS-based observations for studying the Earth’s ionosphere.
More than 4,000 pulsars have now been discovered \citep{mht+05,gpps25}.
Observations of them located at high latitudes and over oceanic regions can effectively fill the gaps in ionospheric measurements in these areas, thereby contributing to the development of a more comprehensive global ionospheric model.
In addition, with the continued development and operation of low-frequency radio telescopes worldwide—such as LOFAR in Europe, the FAST core array \citep{jcg+24}, and the Square Kilometre Array \citep{dhs+09}—the volume of available pulsar observations will continue to grow.
Consequently, large volumes of future pulsar observations are expected to become an important additional data source for ionospheric studies. 
When combined with other datasets, such as GNSS measurements, these pulsar observations may further improve the accuracy of ionospheric reconstruction.
An improvement in ionospheric measurement accuracy will, in turn, enable more precise calibration of the ionospheric RM. 

\begin{acknowledgements}
This work is supported by NSFC grant No.~12503056, grants from Beijing Nova Program  (No.~20250484786), National Science Foundation, China, no. 12421003, the CAS-MPG LEGACY project, the Max-Planck Partner Group, the Strategic Priority Research Program of the Chinese Academy of Sciences (Nos. XDA0350501), and the Major Science and Technology Program of Xinjiang Uygur Autonomous Region, grant No. 2022A03013-2.
J.P.W.V.\ acknowledges support by the Deutsche Forschungsgemeinschaft (DFG) through the Heisenberg programme (Project No.\ 433075039).
MB acknowledges support by the Deutsche Forschungsgemeinschaft (DFG, German Research Foundation) under Germany’s Excellence Strategy – EXC 2121 `Quantum Universe' – 390833306.
RJD acknowledges support from BMFTR ErUM-Pro under grant 05A23PC2. DJS acknowledges support from BMBFT ErUM-Pro under grant 05A23PB1.
LOFAR \citep{vwg+13} is the Low Frequency Array designed and constructed by ASTRON. It has observing, data processing, and data storage facilities in several countries, that are owned by various parties (each with their own funding sources), and that are collectively operated by the ILT foundation under a joint scientific policy. The ILT resources have benefitted from the following recent major funding sources: CNRS-INSU, Observatoire de Paris and Universit\'{e} d'Orl\'{e}ans, France; BMFTR, MIWF-NRW, MPG, Germany; Science Foundation Ireland (SFI), Department of Business, Enterprise and Innovation (DBEI), Ireland; NWO, The Netherlands; The Science and Technology Facilities Council, UK.
This paper uses data obtained with the German LOFAR stations,
during station-owners time.
We made use of data from
the Potsdam (DE604) LOFAR station funded by the
Leibniz-Institut für Astrophysik Potsdam (AIP), Potsdam;
the Jülich (DE605) LOFAR station supported by the
BMFTR Verbundforschung project D-LOFAR I (grant 05A08LJ1);
and the Norderstedt (DE609) LOFAR station funded by the
BMFTR Verbundforschung project D-LOFAR II (grant 05A11LJ1).
The observations of the German LOFAR stations
were carried out in stand-alone GLOW mode,
which is technically operated and supported by
the Max-Planck-Institut für Radioastronomie, the Forschungszentrum
Jülich and Bielefeld University. We acknowledge support and
operation of the GLOW network, computing and storage facilities by
the FZ-Jülich, the MPIfR and Bielefeld University and financial support
from BMFTR D-LOFAR III (grant 05A14PBA) and D-LOFAR IV (grant 05A17PBA),
and by the states of Nordrhein-Westfalia and Hamburg.
\end{acknowledgements}

\bibliography{main}{}

@ARTICLE{pai67,
       author = {{Pacini}, F.},
        title = "{Energy Emission from a Neutron Star}",
      journal = {\nat},
         year = 1967,
        month = nov,
       volume = {216},
       number = {5115},
        pages = {567-568},
          doi = {10.1038/216567a0},
       adsurl = {https://ui.adsabs.harvard.edu/abs/1967Natur.216..567P}
}

@ARTICLE{sch68,
       author = {{Scheuer}, P.~A.~G.},
        title = "{Amplitude Variations in Pulsed Radio Sources}",
      journal = {\nat},
         year = 1968,
        month = jun,
       volume = {218},
       number = {5145},
        pages = {920-922},
          doi = {10.1038/218920a0},
       adsurl = {https://ui.adsabs.harvard.edu/abs/1968Natur.218..920S}
}

@ARTICLE{jcg+24,
       author = {{Jiang}, Peng and {Chen}, Rurong and {Gan}, Hengqian and {Sun}, Jinghai and {Zhu}, Boqin and {Li}, Hui and {Zhu}, Weiwei and {Wu}, Jingwen and {Chen}, Xuelei and {Zhang}, Haiyan and {An}, Tao},
        title = "{The FAST Core Array}",
      journal = {Astronomical Techniques and Instruments},
         year = 2024,
        month = jan,
       volume = {1},
       number = {2},
        pages = {84-94},
          doi = {10.61977/ati2024012},
archivePrefix = {arXiv},
       eprint = {2408.12826},
 primaryClass = {astro-ph.IM},
       adsurl = {https://ui.adsabs.harvard.edu/abs/2024AstTI...1...84J}
}

@ARTICLE{man71,
       author = {{Manchester}, R.~N.},
        title = "{Observations of Pulsar Polarization at 410 and 1665 MHz}",
      journal = {\apjs},
         year = 1971,
        month = sep,
       volume = {23},
        pages = {283},
          doi = {10.1086/190240},
       adsurl = {https://ui.adsabs.harvard.edu/abs/1971ApJS...23..283M}
}

@ARTICLE{ric77,
       author = {{Rickett}, B.~J.},
        title = "{Interstellar scattering and scintillation of radio waves.}",
      journal = {\araa},
         year = 1977,
        month = jan,
       volume = {15},
        pages = {479-504},
          doi = {10.1146/annurev.aa.15.090177.002403},
       adsurl = {https://ui.adsabs.harvard.edu/abs/1977ARA&A..15..479R}
}

@BOOK{lk12,
       author = {{Lorimer}, D.~R. and {Kramer}, M.},
        title = "{Handbook of Pulsar Astronomy}",
         year = 2012,
       adsurl = {https://ui.adsabs.harvard.edu/abs/2012hpa..book.....L}
}

@ARTICLE{khz+25,
       author = {{Kaur}, D. and {Hobbs}, G. and {Zic}, A. and {Dawson}, J.~R. and {Morgan}, J. and {Ling}, W. and {Camtepe}, S. and {Pieprzyk}, J. and {Cheung}, M.~C.~M.},
        title = "{Unlocking the hidden potential of pulsar astronomy}",
      journal = {\na},
         year = 2025,
        month = dec,
       volume = {121},
          eid = {102460},
        pages = {102460},
          doi = {10.1016/j.newast.2025.102460},
archivePrefix = {arXiv},
       eprint = {2506.08056},
 primaryClass = {astro-ph.IM},
       adsurl = {https://ui.adsabs.harvard.edu/abs/2025NewA..12102460K}
}

@ARTICLE{bb05,
       author = {{Brentjens}, M.~A. and {de Bruyn}, A.~G.},
        title = "{Faraday rotation measure synthesis}",
      journal = {\aap},
         year = 2005,
        month = oct,
       volume = {441},
       number = {3},
        pages = {1217-1228},
          doi = {10.1051/0004-6361:20052990},
archivePrefix = {arXiv},
       eprint = {astro-ph/0507349},
 primaryClass = {astro-ph},
       adsurl = {https://ui.adsabs.harvard.edu/abs/2005A&A...441.1217B}
}

@ARTICLE{ssh+13,
       author = {{Sotomayor-Beltran}, C. and {Sobey}, C. and {Hessels}, J.~W.~T. and {de Bruyn}, G. and {Noutsos}, A. and {Alexov}, A. and {Anderson}, J. and {Asgekar}, A. and {Avruch}, I.~M. and {Beck}, R. and {Bell}, M.~E. and {Bell}, M.~R. and {Bentum}, M.~J. and {Bernardi}, G. and {Best}, P. and {Birzan}, L. and {Bonafede}, A. and {Breitling}, F. and {Broderick}, J. and {Brouw}, W.~N. and {Br{\"u}ggen}, M. and {Ciardi}, B. and {de Gasperin}, F. and {Dettmar}, R. -J. and {van Duin}, A. and {Duscha}, S. and {Eisl{\"o}ffel}, J. and {Falcke}, H. and {Fallows}, R.~A. and {Fender}, R. and {Ferrari}, C. and {Frieswijk}, W. and {Garrett}, M.~A. and {Grie{\ss}meier}, J. and {Grit}, T. and {Gunst}, A.~W. and {Hassall}, T.~E. and {Heald}, G. and {Hoeft}, M. and {Horneffer}, A. and {Iacobelli}, M. and {Juette}, E. and {Karastergiou}, A. and {Keane}, E. and {Kohler}, J. and {Kramer}, M. and {Kondratiev}, V.~I. and {Koopmans}, L.~V.~E. and {Kuniyoshi}, M. and {Kuper}, G. and {van Leeuwen}, J. and {Maat}, P. and {Macario}, G. and {Markoff}, S. and {McKean}, J.~P. and {Mulcahy}, D.~D. and {Munk}, H. and {Orru}, E. and {Paas}, H. and {Pandey-Pommier}, M. and {Pilia}, M. and {Pizzo}, R. and {Polatidis}, A.~G. and {Reich}, W. and {R{\"o}ttgering}, H. and {Serylak}, M. and {Sluman}, J. and {Stappers}, B.~W. and {Tagger}, M. and {Tang}, Y. and {Tasse}, C. and {ter Veen}, S. and {Vermeulen}, R. and {van Weeren}, R.~J. and {Wijers}, R.~A.~M.~J. and {Wijnholds}, S.~J. and {Wise}, M.~W. and {Wucknitz}, O. and {Yatawatta}, S. and {Zarka}, P.},
        title = "{Calibrating high-precision Faraday rotation measurements for LOFAR and the next generation of low-frequency radio telescopes}",
      journal = {\aap},
         year = 2013,
        month = apr,
       volume = {552},
          eid = {A58},
        pages = {A58},
          doi = {10.1051/0004-6361/201220728},
archivePrefix = {arXiv},
       eprint = {1303.6230},
 primaryClass = {astro-ph.IM},
       adsurl = {https://ui.adsabs.harvard.edu/abs/2013A&A...552A..58S}
}

@ARTICLE{pnt+19,
       author = {{Porayko}, N.~K. and {Noutsos}, A. and {Tiburzi}, C. and {Verbiest}, J.~P.~W. and {Horneffer}, A. and {K{\"u}nsem{\"o}ller}, J. and {Os{\l}owski}, S. and {Kramer}, M. and {Schnitzeler}, D.~H.~F.~M. and {Anderson}, J.~M. and {Br{\"u}ggen}, M. and {Grie{\ss}meier}, J. -M. and {Hoeft}, M. and {Schwarz}, D.~J. and {Serylak}, M. and {Wucknitz}, O.},
        title = "{Testing the accuracy of the ionospheric Faraday rotation corrections through LOFAR observations of bright northern pulsars}",
      journal = {\mnras},
         year = 2019,
        month = mar,
       volume = {483},
       number = {3},
        pages = {4100-4113},
          doi = {10.1093/mnras/sty3324},
archivePrefix = {arXiv},
       eprint = {1812.01463},
 primaryClass = {astro-ph.IM},
       adsurl = {https://ui.adsabs.harvard.edu/abs/2019MNRAS.483.4100P}
}

@ARTICLE{vwg+13,
       author = {{van Haarlem}, M.~P. and {Wise}, M.~W. and {Gunst}, A.~W. and {Heald}, G. and {McKean}, J.~P. and {Hessels}, J.~W.~T. and {de Bruyn}, A.~G. and {Nijboer}, R. and {Swinbank}, J. and {Fallows}, R. and {Brentjens}, M. and {Nelles}, A. and {Beck}, R. and {Falcke}, H. and {Fender}, R. and {H{\"o}randel}, J. and {Koopmans}, L.~V.~E. and {Mann}, G. and {Miley}, G. and {R{\"o}ttgering}, H. and {Stappers}, B.~W. and {Wijers}, R.~A.~M.~J. and {Zaroubi}, S. and {van den Akker}, M. and {Alexov}, A. and {Anderson}, J. and {Anderson}, K. and {van Ardenne}, A. and {Arts}, M. and {Asgekar}, A. and {Avruch}, I.~M. and {Batejat}, F. and {B{\"a}hren}, L. and {Bell}, M.~E. and {Bell}, M.~R. and {van Bemmel}, I. and {Bennema}, P. and {Bentum}, M.~J. and {Bernardi}, G. and {Best}, P. and {B{\^\i}rzan}, L. and {Bonafede}, A. and {Boonstra}, A. -J. and {Braun}, R. and {Bregman}, J. and {Breitling}, F. and {van de Brink}, R.~H. and {Broderick}, J. and {Broekema}, P.~C. and {Brouw}, W.~N. and {Br{\"u}ggen}, M. and {Butcher}, H.~R. and {van Cappellen}, W. and {Ciardi}, B. and {Coenen}, T. and {Conway}, J. and {Coolen}, A. and {Corstanje}, A. and {Damstra}, S. and {Davies}, O. and {Deller}, A.~T. and {Dettmar}, R. -J. and {van Diepen}, G. and {Dijkstra}, K. and {Donker}, P. and {Doorduin}, A. and {Dromer}, J. and {Drost}, M. and {van Duin}, A. and {Eisl{\"o}ffel}, J. and {van Enst}, J. and {Ferrari}, C. and {Frieswijk}, W. and {Gankema}, H. and {Garrett}, M.~A. and {de Gasperin}, F. and {Gerbers}, M. and {de Geus}, E. and {Grie{\ss}meier}, J. -M. and {Grit}, T. and {Gruppen}, P. and {Hamaker}, J.~P. and {Hassall}, T. and {Hoeft}, M. and {Holties}, H.~A. and {Horneffer}, A. and {van der Horst}, A. and {van Houwelingen}, A. and {Huijgen}, A. and {Iacobelli}, M. and {Intema}, H. and {Jackson}, N. and {Jelic}, V. and {de Jong}, A. and {Juette}, E. and {Kant}, D. and {Karastergiou}, A. and {Koers}, A. and {Kollen}, H. and {Kondratiev}, V.~I. and {Kooistra}, E. and {Koopman}, Y. and {Koster}, A. and {Kuniyoshi}, M. and {Kramer}, M. and {Kuper}, G. and {Lambropoulos}, P. and {Law}, C. and {van Leeuwen}, J. and {Lemaitre}, J. and {Loose}, M. and {Maat}, P. and {Macario}, G. and {Markoff}, S. and {Masters}, J. and {McFadden}, R.~A. and {McKay-Bukowski}, D. and {Meijering}, H. and {Meulman}, H. and {Mevius}, M. and {Middelberg}, E. and {Millenaar}, R. and {Miller-Jones}, J.~C.~A. and {Mohan}, R.~N. and {Mol}, J.~D. and {Morawietz}, J. and {Morganti}, R. and {Mulcahy}, D.~D. and {Mulder}, E. and {Munk}, H. and {Nieuwenhuis}, L. and {van Nieuwpoort}, R. and {Noordam}, J.~E. and {Norden}, M. and {Noutsos}, A. and {Offringa}, A.~R. and {Olofsson}, H. and {Omar}, A. and {Orr{\'u}}, E. and {Overeem}, R. and {Paas}, H. and {Pandey-Pommier}, M. and {Pandey}, V.~N. and {Pizzo}, R. and {Polatidis}, A. and {Rafferty}, D. and {Rawlings}, S. and {Reich}, W. and {de Reijer}, J. -P. and {Reitsma}, J. and {Renting}, G.~A. and {Riemers}, P. and {Rol}, E. and {Romein}, J.~W. and {Roosjen}, J. and {Ruiter}, M. and {Scaife}, A. and {van der Schaaf}, K. and {Scheers}, B. and {Schellart}, P. and {Schoenmakers}, A. and {Schoonderbeek}, G. and {Serylak}, M. and {Shulevski}, A. and {Sluman}, J. and {Smirnov}, O. and {Sobey}, C. and {Spreeuw}, H. and {Steinmetz}, M. and {Sterks}, C.~G.~M. and {Stiepel}, H. -J. and {Stuurwold}, K. and {Tagger}, M. and {Tang}, Y. and {Tasse}, C. and {Thomas}, I. and {Thoudam}, S. and {Toribio}, M.~C. and {van der Tol}, B. and {Usov}, O. and {van Veelen}, M. and {van der Veen}, A. -J. and {ter Veen}, S. and {Verbiest}, J.~P.~W. and {Vermeulen}, R. and {Vermaas}, N. and {Vocks}, C. and {Vogt}, C. and {de Vos}, M. and {van der Wal}, E. and {van Weeren}, R. and {Weggemans}, H. and {Weltevrede}, P. and {White}, S. and {Wijnholds}, S.~J. and {Wilhelmsson}, T. and {Wucknitz}, O. and {Yatawatta}, S. and {Zarka}, P. and {Zensus}, A.},
        title = "{LOFAR: The LOw-Frequency ARray}",
      journal = {\aap},
         year = 2013,
        month = aug,
       volume = {556},
          eid = {A2},
        pages = {A2},
          doi = {10.1051/0004-6361/201220873},
archivePrefix = {arXiv},
       eprint = {1305.3550},
 primaryClass = {astro-ph.IM},
       adsurl = {https://ui.adsabs.harvard.edu/abs/2013A&A...556A...2V}
}

@ARTICLE{tvs+19,
       author = {{Tiburzi}, C. and {Verbiest}, J.~P.~W. and {Shaifullah}, G.~M. and {Janssen}, G.~H. and {Anderson}, J.~M. and {Horneffer}, A. and {K{\"u}nsem{\"o}ller}, J. and {Os{\l}owski}, S. and {Donner}, J.~Y. and {Kramer}, M. and {Kumari}, A. and {Porayko}, N.~K. and {Zucca}, P. and {Ciardi}, B. and {Dettmar}, R. -J. and {Grie{\ss}meier}, J. -M. and {Hoeft}, M. and {Serylak}, M.},
        title = "{On the usefulness of existing solar wind models for pulsar timing corrections}",
      journal = {\mnras},
         year = 2019,
        month = jul,
       volume = {487},
       number = {1},
        pages = {394-408},
          doi = {10.1093/mnras/stz1278},
archivePrefix = {arXiv},
       eprint = {1905.02989},
 primaryClass = {astro-ph.HE},
       adsurl = {https://ui.adsabs.harvard.edu/abs/2019MNRAS.487..394T}
}

@ARTICLE{dvt+20,
       author = {{Donner}, J.~Y. and {Verbiest}, J.~P.~W. and {Tiburzi}, C. and {Os{\l}owski}, S. and {K{\"u}nsem{\"o}ller}, J. and {Bak Nielsen}, A.-S. and {Grie{\ss}meier}, J.-M. and {Serylak}, M. and {Kramer}, M. and {Anderson}, J.~M. and {Wucknitz}, O. and {Keane}, E. and {Kondratiev}, V. and {Sobey}, C. and {McKee}, J.~W. and {Bilous}, A.~V. and {Breton}, R.~P. and {Br{\"u}ggen}, M. and {Ciardi}, B. and {Hoeft}, M. and {van Leeuwen}, J. and {Vocks}, C.},
        title = "{Dispersion measure variability for 36 millisecond pulsars at 150 MHz with LOFAR}",
      journal = {\aap},
         year = 2020,
        month = dec,
       volume = {644},
          eid = {A153},
        pages = {A153},
          doi = {10.1051/0004-6361/202039517},
archivePrefix = {arXiv},
       eprint = {2011.13742},
 primaryClass = {astro-ph.HE},
       adsurl = {https://ui.adsabs.harvard.edu/abs/2020A&A...644A.153D}
}

@ARTICLE{vdo12,
       author = {{van Straten}, Willem and {Demorest}, Paul and {Oslowski}, Stefan},
        title = "{Pulsar Data Analysis with PSRCHIVE}",
      journal = {Astronomical Research and Technology},
         year = 2012,
        month = jul,
       volume = {9},
       number = {3},
        pages = {237-256},
          doi = {10.48550/arXiv.1205.6276},
archivePrefix = {arXiv},
       eprint = {1205.6276},
 primaryClass = {astro-ph.IM},
       adsurl = {https://ui.adsabs.harvard.edu/abs/2012AR&T....9..237V}
}

@ARTICLE{lkg+16,
       author = {{Lazarus}, P. and {Karuppusamy}, R. and {Graikou}, E. and {Caballero}, R.~N. and {Champion}, D.~J. and {Lee}, K.~J. and {Verbiest}, J.~P.~W. and {Kramer}, M.},
        title = "{Prospects for high-precision pulsar timing with the new Effelsberg PSRIX backend}",
      journal = {\mnras},
         year = 2016,
        month = may,
       volume = {458},
       number = {1},
        pages = {868-880},
          doi = {10.1093/mnras/stw189},
archivePrefix = {arXiv},
       eprint = {1601.06194},
 primaryClass = {astro-ph.IM},
       adsurl = {https://ui.adsabs.harvard.edu/abs/2016MNRAS.458..868L}
}

@ARTICLE{gpps25,
       author = {{Han}, J.~L. and {Zhou}, D.~J. and {Wang}, C. and {Su}, W.~Q. and {Yan}, Yi and {Jing}, W.~C. and {Yang}, Z.~L. and {Wang}, P.~F. and {Wang}, T. and {Xu}, J. and {Cai}, N.~N. and {Sun}, J.~H. and {Yang}, Q.~L. and {Xu}, R.~X. and {Wang}, H.~G. and {You}, X.~P.},
        title = "{The FAST Galactic Plane Pulsar Snapshot Survey. VI. The Discovery of 473 New Pulsars}",
      journal = {Research in Astronomy and Astrophysics},
         year = 2025,
        month = jan,
       volume = {25},
       number = {1},
          eid = {014001},
        pages = {014001},
          doi = {10.1088/1674-4527/ada3b7},
archivePrefix = {arXiv},
       eprint = {2411.15961},
 primaryClass = {astro-ph.HE},
       adsurl = {https://ui.adsabs.harvard.edu/abs/2025RAA....25a4001H}
}

@ARTICLE{dhs+09,
       author = {{Dewdney}, P.~E. and {Hall}, P.~J. and {Schilizzi}, R.~T. and {Lazio}, T.~J.~L.~W.},
        title = "{The Square Kilometre Array}",
      journal = {IEEE Proceedings},
         year = 2009,
        month = aug,
       volume = {97},
       number = {8},
        pages = {1482-1496},
          doi = {10.1109/JPROC.2009.2021005},
       adsurl = {https://ui.adsabs.harvard.edu/abs/2009IEEEP..97.1482D}
}

@ARTICLE{ymh+11,
       author = {{Yan}, W.~M. and {Manchester}, R.~N. and {Hobbs}, G. and {van Straten}, W. and {Reynolds}, J.~E. and {Wang}, N. and {Bailes}, M. and {Bhat}, N.~D.~R. and {Burke-Spolaor}, S. and {Champion}, D.~J. and {Chaudhary}, A. and {Coles}, W.~A. and {Hotan}, A.~W. and {Khoo}, J. and {Oslowski}, S. and {Sarkissian}, J.~M. and {Yardley}, D.~R.~B.},
        title = "{Rotation measure variations for 20 millisecond pulsars}",
      journal = {\apss},
         year = 2011,
        month = oct,
       volume = {335},
       number = {2},
        pages = {485-498},
          doi = {10.1007/s10509-011-0756-0},
archivePrefix = {arXiv},
       eprint = {1105.4213},
 primaryClass = {astro-ph.SR},
       adsurl = {https://ui.adsabs.harvard.edu/abs/2011Ap&SS.335..485Y}
}

@article{sja+24,
  title={Mapping the ionosphere with millions of phones},
  author={Smith, Jamie and Kast, Anton and Geraschenko, Anton and Morton, Y Jade and Brenner, Michael P and van Diggelen, Frank and Williams, Brian P},
  journal={Nature},
  volume={635},
  number={8038},
  pages={365--369},
  year={2024},
  doi = {10.1038/s41586-024-08072-x},
  publisher={Nature Publishing Group UK London}
}

@article{mwy+98,
  title={A global mapping technique for GPS-derived ionospheric total electron content measurements},
  author={Mannucci, AJ and Wilson, BD and Yuan, DN and Ho, CH and Lindqwister, UJ and Runge, TF},
  journal={Radio science},
  volume={33},
  number={3},
  pages={565--582},
  year={1998},
  publisher={AGU}
}

@article{fel03,
title = {The International GPS Service (IGS) Ionosphere Working Group},
journal = {Advances in Space Research},
volume = {31},
number = {3},
pages = {635-644},
year = {2003},
note = {Description of the Low Latitude and Equatorial Ionosphere in the International Reference Ionosphere},
issn = {0273-1177},
doi = {https://doi.org/10.1016/S0273-1177(03)00029-2},
url = {https://www.sciencedirect.com/science/article/pii/S0273117703000292},
author = {J. Feltens},
}

@ARTICLE{usm+13,
       author = {{Ulyanov}, O.~M. and {Shevtsova}, A.~I. and {Mukha}, D.~V. and {Seredkina}, A.~A.},
        title = "{Investigation of the Earth ionosphere using the radio emission of pulsars}",
      journal = {Baltic Astronomy},
         year = 2013,
        month = jan,
       volume = {22},
        pages = {53-65},
          doi = {10.1515/astro-2017-0147},
archivePrefix = {arXiv},
       eprint = {1302.3821},
 primaryClass = {astro-ph.EP},
       adsurl = {https://ui.adsabs.harvard.edu/abs/2013BaltA..22...53U}
}

@ARTICLE{mht+05,
       author = {{Manchester}, R.~N. and {Hobbs}, G.~B. and {Teoh}, A. and {Hobbs}, M.},
        title = "{The Australia Telescope National Facility Pulsar Catalogue}",
      journal = {\aj},
         year = 2005,
        month = apr,
       volume = {129},
       number = {4},
        pages = {1993-2006},
          doi = {10.1086/428488},
archivePrefix = {arXiv},
       eprint = {astro-ph/0412641},
 primaryClass = {astro-ph},
       adsurl = {https://ui.adsabs.harvard.edu/abs/2005AJ....129.1993M}
}

@ARTICLE{nsk+15,
       author = {{Noutsos}, A. and {Sobey}, C. and {Kondratiev}, V.~I. and {Weltevrede}, P. and {Verbiest}, J.~P.~W. and {Karastergiou}, A. and {Kramer}, M. and {Kuniyoshi}, M. and {Alexov}, A. and {Breton}, R.~P. and {Bilous}, A.~V. and {Cooper}, S. and {Falcke}, H. and {Grie{\ss}meier}, J.-M. and {Hassall}, T.~E. and {Hessels}, J.~W.~T. and {Keane}, E.~F. and {Os{\l}owski}, S. and {Pilia}, M. and {Serylak}, M. and {Stappers}, B.~W. and {ter Veen}, S. and {van Leeuwen}, J. and {Zagkouris}, K. and {Anderson}, K. and {B{\"a}hren}, L. and {Bell}, M. and {Broderick}, J. and {Carbone}, D. and {Cendes}, Y. and {Coenen}, T. and {Corbel}, S. and {Eisl{\"o}ffel}, J. and {Fender}, R. and {Garsden}, H. and {Jonker}, P. and {Law}, C. and {Markoff}, S. and {Masters}, J. and {Miller-Jones}, J. and {Molenaar}, G. and {Osten}, R. and {Pietka}, M. and {Rol}, E. and {Rowlinson}, A. and {Scheers}, B. and {Spreeuw}, H. and {Staley}, T. and {Stewart}, A. and {Swinbank}, J. and {Wijers}, R. and {Wijnands}, R. and {Wise}, M. and {Zarka}, P. and {van der Horst}, A.},
        title = "{Pulsar polarisation below 200 MHz: Average profiles and propagation effects}",
      journal = {\aap},
         year = 2015,
        month = apr,
       volume = {576},
          eid = {A62},
        pages = {A62},
          doi = {10.1051/0004-6361/201425186},
archivePrefix = {arXiv},
       eprint = {1501.03312},
 primaryClass = {astro-ph.GA},
       adsurl = {https://ui.adsabs.harvard.edu/abs/2015A&A...576A..62N}
}

@article{yl+21,
  title = {Feasibility analysis of GNSS-based ionospheric TEC estimation in low-latitude regions},
  author = {{Yu}, S.~W. and {Liu}, Z.~Z.},
  journal = {Satellite Navigation},
  year = {2021},
  volume = {2},
  number = {1},
  pages = {20},
  doi = {10.1186/s43020-021-00051-1},
}

@ARTICLE{lbr+23,
       author = {{Li}, Dongzi and {Bilous}, Anna and {Ransom}, Scott and {Main}, Robert and {Yang}, Yuan-Pei},
        title = "{A highly magnetized environment in a pulsar binary system}",
      journal = {\nat},
         year = 2023,
        month = jun,
       volume = {618},
       number = {7965},
        pages = {484-488},
          doi = {10.1038/s41586-023-05983-z},
archivePrefix = {arXiv},
       eprint = {2205.07917},
 primaryClass = {astro-ph.HE},
       adsurl = {https://ui.adsabs.harvard.edu/abs/2023Natur.618..484L}
}

@article{zhl21,
  author  = {Zhao, Jiaojiao and Hern{\'a}ndez-Pajares, Manuel and Li, Zishen and Wang, Ningbo and Yuan, Hong},
  title   = {Integrity investigation of global ionospheric TEC maps for high-precision positioning},
  journal = {Journal of Geodesy},
  year    = {2021},
  volume  = {95},
  number  = {3},
  pages   = {35},
  doi     = {10.1007/s00190-021-01487-8},
  url     = {https://doi.org/10.1007/s00190-021-01487-8},
  issn    = {1432-1394}
}

@ARTICLE{zya+20,
       author = {{Zhuravlev}, Vladimir I. and {Yermolaev}, Yu. I. and {Andrianov}, A.~S.},
        title = "{Probing the ionosphere by the pulsar B0950+08 with help of RadioAstron ground-space baselines}",
      journal = {\mnras},
         year = 2020,
        month = feb,
       volume = {491},
       number = {4},
        pages = {5843-5851},
          doi = {10.1093/mnras/stz3370},
archivePrefix = {arXiv},
       eprint = {1906.10435},
 primaryClass = {astro-ph.IM},
       adsurl = {https://ui.adsabs.harvard.edu/abs/2020MNRAS.491.5843Z}
}

@ARTICLE{bp24,
       author = {{Burgin}, M.~S. and {Popov}, M.~V.},
        title = "{Probing the Ionosphere with Pulses from the Pulsar B2016+28 at a Frequency of 324 MHz}",
      journal = {Astronomy Reports},
         year = 2024,
        month = mar,
       volume = {68},
       number = {3},
        pages = {257-267},
          doi = {10.1134/S1063772924700276},
       adsurl = {https://ui.adsabs.harvard.edu/abs/2024ARep...68..257B}
}

@ARTICLE{hzw+25,
       author = {{He}, Lina and {Zhu}, Qinghao and {Wang}, Cheng},
        title = "{Analysis of the global spatiotemporal characteristics of ionospheric noontime bite-outs}",
      journal = {Satellite Navigation},
         year = 2025,
        month = dec,
       volume = {6},
       number = {1},
          eid = {11},
        pages = {11},
          doi = {10.1186/s43020-025-00164-x},
       adsurl = {https://ui.adsabs.harvard.edu/abs/2025SatNa...6...11H}
}

@ARTICLE{jwy+23,
       author = {{Jiang}, Chunhua and {Wang}, Wenbin and {Yang}, Guobin and {Zhao}, Zhengyu},
        title = "{A numerical study of noontime bite-outs in F2-region electron density}",
      journal = {Advances in Space Research},
         year = 2023,
        month = feb,
       volume = {71},
       number = {3},
        pages = {1818-1826},
          doi = {10.1016/j.asr.2022.09.042},
       adsurl = {https://ui.adsabs.harvard.edu/abs/2023AdSpR..71.1818J}
}

@article{lgr+14,
author = {Lynn, Kenneth J. W. and Gardiner-Garden, Robert S. and Heitmann, Andrew},
title = {The spatial and temporal structure of twin peaks and midday bite out in foF2 (with associated height changes) in the Australian and South Pacific low midlatitude ionosphere},
journal = {Journal of Geophysical Research: Space Physics},
volume = {119},
number = {12},
pages = {10,294-10,304},
doi = {https://doi.org/10.1002/2014JA020617},
url = {https://agupubs.onlinelibrary.wiley.com/doi/abs/10.1002/2014JA020617},
year = {2014}
}

@ARTICLE{app+46,
       author = {{Appleton}, Edward V.},
        title = "{Two Anomalies in the Ionosphere}",
      journal = {\nat},
         year = 1946,
        month = may,
       volume = {157},
       number = {3995},
        pages = {691},
          doi = {10.1038/157691a0},
       adsurl = {https://ui.adsabs.harvard.edu/abs/1946Natur.157..691A}
}

@ARTICLE{lia+47,
       author = {{Liang}, P.~H.},
        title = "{F$_{2}$ Ionization and Geomagnetic Latitudes}",
      journal = {\nat},
         year = 1947,
        month = nov,
       volume = {160},
       number = {4071},
        pages = {642-643},
          doi = {10.1038/160642a0},
       adsurl = {https://ui.adsabs.harvard.edu/abs/1947Natur.160..642L}
}

@software{mev+18,
       author = {{Mevius}, Maaijke},
        title = "{RMextract: Ionospheric Faraday Rotation calculator}",
 howpublished = {Astrophysics Source Code Library, record ascl:1806.024},
         year = 2018,
        month = jun,
          eid = {ascl:1806.024},
archivePrefix = {ascl},
       eprint = {1806.024},
       adsurl = {https://ui.adsabs.harvard.edu/abs/2018ascl.soft06024M}
}

@misc{wmm2020,
  author       = {{NCEI Geomagnetic Modeling Team} and {British Geological Survey}},
  title        = {World Magnetic Model 2020},
  year         = {2020},
  publisher    = {NOAA National Centers for Environmental Information},
  doi          = {10.25921/11v3-da71},
  url          = {https://doi.org/10.25921/11v3-da71}
}

@ARTICLE{nfc+25,
       author = {{Nair}, Manoj and {Fillion}, Martin and {Chulliat}, Arnaud and {Califf}, Sam},
        title = "{Global Geomagnetic Model Errors as a Function of Altitude and Geomagnetic Activity}",
      journal = {Space Weather},
         year = 2025,
        month = oct,
       volume = {23},
       number = {10},
          eid = {e2025SW004579},
        pages = {e2025SW004579},
          doi = {10.1029/2025SW004579},
       adsurl = {https://ui.adsabs.harvard.edu/abs/2025SpWea..2304579N}
}

@ARTICLE{lzw+25,
       author = {{Li}, Zhiyao and {Zhong}, Jiahao and {Wang}, Ningbo and {Hao}, Yongqiang and {Wan}, Xin and {Jakowski}, N. and {Hoque}, M.~M. and {Liu}, Ang and {Li}, Zishen and {Liu}, Runqing and {Nykiel}, Grzegorz},
        title = "{Ionospheric gradient estimation using ground-based GEO observations for monitoring multi-scale ionospheric dynamics}",
      journal = {Satellite Navigation},
         year = 2025,
        month = dec,
       volume = {6},
       number = {1},
          eid = {30},
        pages = {30},
          doi = {10.1186/s43020-025-00187-4},
       adsurl = {https://ui.adsabs.harvard.edu/abs/2025SatNa...6...30L}
}

@ARTICLE{pbg+26,
       author = {{Perley}, Richard A. and {Butler}, Bryan J. and {Greisen}, Eric W. and {Hugo}, Benjamin V. and {Tremou}, Evangelia and {Willis}, A.~G.},
        title = "{Correcting Ionospheric Faraday Rotation for the VLA and MeerKAT}",
      journal = {\apjs},
         year = 2026,
        month = apr,
       volume = {283},
       number = {2},
          eid = {82},
        pages = {82},
          doi = {10.3847/1538-4365/ae503c},
archivePrefix = {arXiv},
       eprint = {2603.09001},
 primaryClass = {astro-ph.IM},
       adsurl = {https://ui.adsabs.harvard.edu/abs/2026ApJS..283...82P}
}

@article{Liu2026,
  author  = {Liu, Ang and Wang, Ningbo and Li, Zishen and Brack, Andreas and Hern{\'a}ndez-Pajares, Manuel and Krankowski, Andrzej and Gini, Francesco and Ghoddousi-Fard, Reza and Liu, Qi and Wang, Liang and Fro{\'n}, Adam and Li, Ang and Liu, Runqing and Xie, Mingqiang},
  title   = {{FAIRS}: A Factor-Adjusted Ionospheric Residual Statistics Approach for Global Ionospheric {RMS} Mapping},
  journal = {Satellite Navigation},
  year    = {2026},
  volume  = {7},
  number  = {1},
  pages   = {20},
  doi     = {10.1186/s43020-026-00209-9},
  url     = {https://doi.org/10.1186/s43020-026-00209-9},
  issn    = {2662-1363}
}
\bibliographystyle{aasjournalv7}

\end{document}